\documentclass[12pt]{article}    
\usepackage{epsfig} 
\newcommand{\beeq}{\begin{eqnarray}}    
\newcommand{\eeeq}{\end{eqnarray}}    
\newcommand{\be}{\begin{equation}}    
\newcommand{\ee}{\end{equation}}    
\newcommand{\asb}{{\bar{\alpha}_s}} 
\begin{document}   
\titlepage    
\begin{flushright}
DFF 383/3/02 \\
March 2002
\end{flushright}    
\vspace*{1in}    
\begin{center}    
{\Large \bf Geometric scaling and QCD evolution}\\    
\vspace*{0.4in}    
J. \ Kwieci\'nski$^{(a)}$    
and A.\ M. \ Sta\'sto$^{(a,b)}$ \\    
\vspace*{0.5cm}    
  
$^{(a)}${\it H. Niewodnicza\'nski Institute of Nuclear Physics,    
 Krak\'ow, Poland} \\    
\vskip 2mm  
$^{(b)}$ {\it  INFN Sezione di Firenze, Via G. Sansone 1, 50019 Sesto Fiorentino (FI), Italy} \\   
\vskip 2mm    
\end{center}    
\vspace*{1cm}    
\centerline{(\today)}    
    
\vskip1cm    
\begin{abstract}    
We study the impact of the QCD DGLAP evolution on the geometric scaling of the   
gluon distributions which is expected to hold at small $x$ within the saturation models.    
To this aim we solve the DGLAP evolution equations with the initial conditions   
provided along the critical line $Q^2=Q_s^2(x)$ with $Q_s^2(x) \sim x^{-\lambda}$ and   
satisfying geometric scaling.  Both fixed and running coupling cases are studied.    
We show that in the fixed coupling case the geometric scaling at low $x$    
is stable against the DGLAP evolution for sufficiently large values of the parameter   
$\lambda$ and   in the double logarithmic approximation of the DGLAP evolution   
this happens for $\lambda \ge 4N_c\alpha_s/\pi$. In the running coupling case geometric scaling is   
found  to be approximately preserved at very small $x$.
 The residual geometric scaling violation in this case can be approximately factored out
and the corresponding form-factor controlling this violation is found.
  
\end{abstract}

\newpage        
\section{Introduction}  
Perturbative QCD predicts very strong power-law rise of the gluon density $xg(x,Q^2)$ in the limit   
$x \rightarrow 0$,   
where, as usual,   $x$ denotes the momentum fraction carried by the gluon and $Q^2$ is the scale at which  
the distribution is probed. 
  This strong rise can eventually violate   
unitarity and so it has to be tamed by  screening effects.  
Those screening effects are provided by multiple parton interactions which lead to the  
non-linear terms in the (BFKL and/or DGLAP) equations \cite{GLR} -   
\cite{WEIGERT}. 
These non-linear terms  reduce the growth of gluon distributions and generate instead the   
parton saturation at sufficiently small values of $x$ and/or $Q^2$  
\cite{GLR} - \cite{GBMS}.\\ 
  
Increase of the gluon distribution and  emergence of the  saturation effects 
imply similar properties   
of the measurable quantities which are driven by the gluon, like the deep inelastic structure function   
$F_2(x,Q^2)$.  This can be most clearly seen in the dipole picture of deep inelastic   
scattering in which the virtual photon - proton total cross section   
$\sigma_{\gamma^*p}(x,Q^2)$ ( 
$\sigma_{\gamma^*p}(x,Q^2) \sim F_2(x,Q^2)/Q^2)$)  
is linked with the  cross section $\sigma_{dp}(x,r)$   
describing the interaction of the   
 $q \bar q$ colour dipole with the proton, where $r$ denotes the transverse size of the   
dipole \cite{MUELLER,NIKO,KGBMW,MANCH}. The dipole-proton cross section is determined by   
the gluon distribution in the proton and in leading order approximation we just have   
$\sigma_{dp}(x,r) \sim \alpha_s(1/r^2)r^2xg(x,1/r^2)$.  Increase and/or saturation  
 of the gluon distribution   
in the small $x$ limit implies similar increase and/or saturation  of the dipole-proton cross section   
and of the cross-section $\sigma_{\gamma^*p}(x,Q^2)$.\\  
 
The successful description of all inclusive and diffractive deep inelastic  
data at HERA by the saturation model \cite{KGBMW} ( see also \cite{BGBK} and \cite{SMUNIER}) suggests that the screening effects might become important in the energy regime probed by the present colliders.  
Important property of the dipole cross section which holds in this model  is   
its geometric scaling, i.e. dependence upon single variable $\tau=r^2Q_s^2(x)$ where   
$Q_s(x)$ is the saturation scale.  This leads to the geometric scaling of     
$\sigma_{\gamma^*p}(x,Q^2)$ itself, i.e. $\sigma_{\gamma^*p}(x,Q^2)=f(Q^2/Q_s^2(x))$    
which is well supported by the experimental data from HERA   
\cite{GEOMETRIC}.  Geometric scaling of the dipole cross-section should imply similar  
scaling of the quantity $\alpha_s(Q^2)xg(x,Q^2)/Q^2$.    
This type of scaling is also found to be an intrinsic property of the non-linear evolution equations \cite{BARLEV,VENUGOPALAN}, \cite{BAL}  - \cite{GBMS}. It turns out that for the equations of type 
\be 
\frac{\partial \phi(x,k)}{\partial \ln(1/x)} \, = \, \asb K \otimes \phi - \asb \phi^2(x,k)\; , \hspace{1cm} (\asb \equiv \frac{N_c \alpha_s}{\pi}) \; ,
\ee 
where $K$ is a linear evolution kernel (for example of BFKL type) there 
exist a region in $x$ and $k$ space such that  
\be 
\phi(x,k) = \phi(Q_s^2(x)/k^2) \hspace{0.5cm} {\rm for} \hspace{0.5cm}  k^2 < Q_s^2(x) \; .
\ee 
For example in the case of the Balitsky-Kovchegov \cite{BAL,KOV1} equation, where $K$ is the BFKL kernel,  
the saturation scale $Q_s^2(x)$ has been found to have a general  power like dependence 
on $x$ , $Q_s^2(x) = Q_0^2 x^{-\lambda}$. The coefficient $\lambda$ which is approximately equal to $4\asb$ in this case, is then a universal quantity and does not depend on the initial conditions for the evolution \cite{LEVLUB} - \cite{GBMS}.

\begin{figure}[t]    
  \vspace*{0.0cm}    
     \centerline{    
         \epsfig{figure=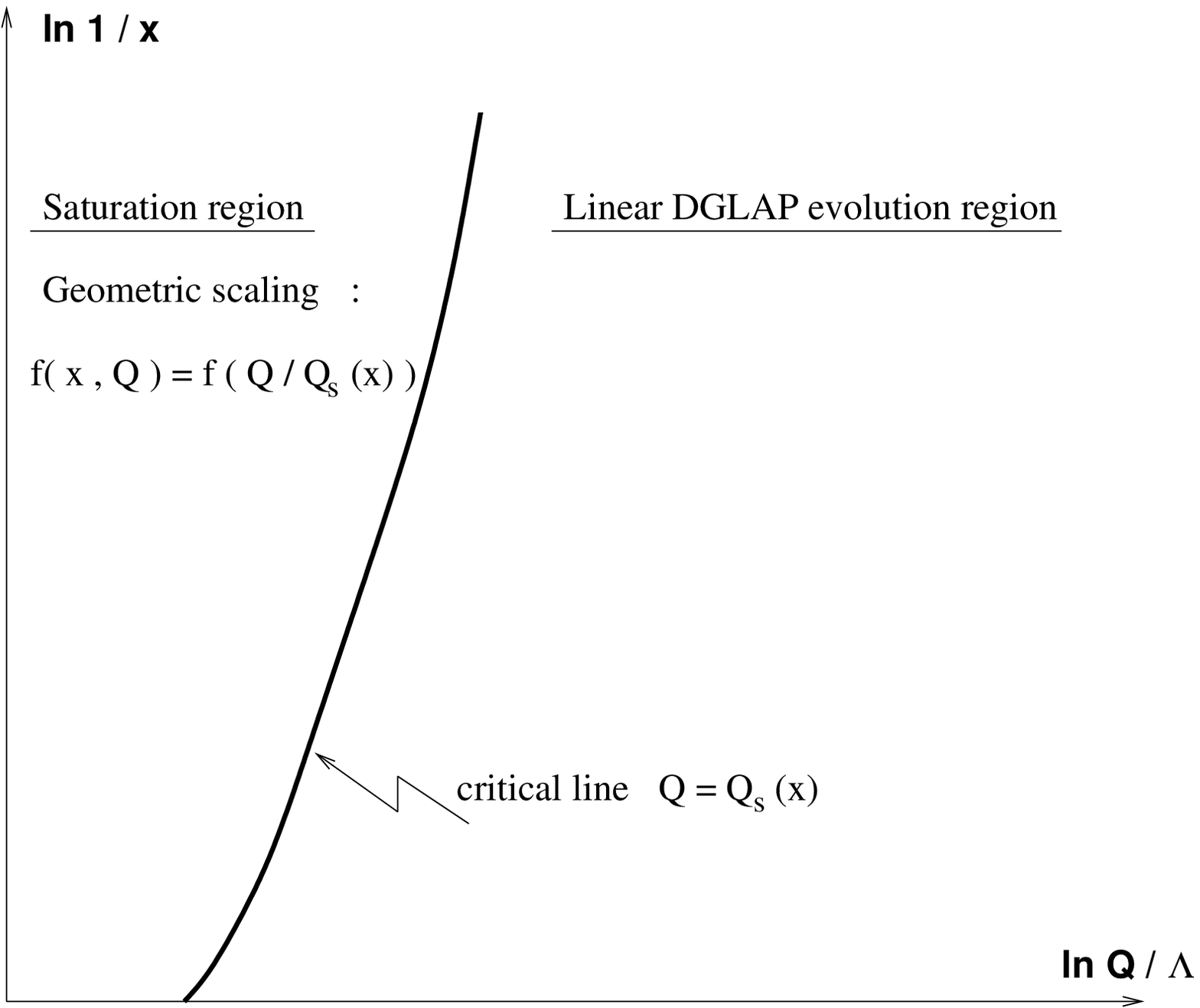,width=15cm}    
           }    
\vspace*{0.5cm}    
\caption{\it Phase diagram in  $(\ln 1/x,\ln Q/\Lambda)$  space. 
Thick line is a critical line 
\mbox{$Q^2 = Q_s^2(x)$} which divides the saturation - scaling regime (to the left) and the linear - DGLAP regime (to the right). 
\label{fig:1}}    
\end{figure}

The main purpose of this paper is to analyse possible compatibility of this scaling with the   
DGLAP evolution equations.  It is expected that the non-linear shadowing effects should   
be weak in the region 'to the right'  of the critical line   
defined by the saturation scale $Q_s^2(x)$ , i.e. for $Q^2 > Q_s^2(x)$, see Fig.~\ref{fig:1}.    
In order to study possible impact of the DGLAP evolution we shall therefore assume the   
geometric scaling parametrisation along the critical line and inspect   
the structure of the solution of the DGLAP equation with those initial conditions.    
 This way of  providing the initial conditions along the critical 
line $Q^2=Q_s^2(x)$ rather than at 
$Q^2=Q_0^2$ with $x$ independent reference scale $Q_0^2$ is the characteristic  feature of the 
saturation effects \cite{GLR}.

The content of our paper is as follows.  In the next section we give semianalytical insight   
into the solution of the DGLAP equation with the starting distributions provided along   
the critical line.   
We study separately the fixed and running coupling cases.   
In Section 3  
we present numerical analysis of our solutions and finally in Section 4 we give  
our conclusions.   
                     
          
\section{Solution of the DGLAP equations from the starting distributions provided along the critical line}  
We wish to understand possible effects of the DGLAP evolution on the   
geometric scaling at low $x$.  This scaling means that certain quantities controlling  
deep inelastic scattering at low $x$, like the dipole-proton cross section  
$\sigma_{dp}(x,r=1/Q)$ or the virtual photon-proton cross section  
$\sigma_{\gamma^*p}$, which are in principle functions of two variables, depend upon  
 the single variable $Q/Q_s(x)$.  The saturation scale $Q_s(x)$, which also specifies the critical  
line increases with decreasing $x$ \\   

\begin {equation}  
Q_s^2(x)=Q_0^2 x^{-\lambda} \; . 
\label{qs}  
\end{equation}  
Let us assume that:     
\begin{enumerate}  
\item For $Q^2<Q_s^2(x)$ the linear evolution is strongly perturbed by the nonlinear effects which  
generate geometric scaling for the dipole cross section 
 \mbox{$\sigma_{dp}(x,r=1/Q)$} and for the related quantities.  
 
\item Geometric scaling for the dipole cross-section implies geometric scaling for  \\ 
$\alpha_s(Q^2) xg(x,Q^2)/Q^2$, where $g(x,Q^2)$ denotes the gluon distribution.  
 This  follows from the LO relation between the dipole cross section   
and the  gluon distribution, i.e.  
$\sigma(x,r^2) \sim r^2 \alpha_s(1/r^2)xg(x,1/r^2)$.    
\item  Geometric scaling for $\alpha_s(Q^2) xg(x,Q^2)/Q^2$  
holds at the boundary $Q^2=Q_s^2(x)$. 
\item For $Q^2 > Q_s^2(x)$  the non-linear screening effects can be neglected and  
evolution of parton densities is governed by the DGLAP equations.     
\end{enumerate}   
We wish to study possible effects of the DGLAP evolution upon the geometric scaling in the region  
$Q^2>Q_s^2(x)$  after  solving   the linear DGLAP evolution equations starting from the  
gluon distribution satisfying this scaling   and  
defined along the critical line $Q_s^2(x)$ (see point 3 above).  We shall discuss  
the fixed and running coupling cases separately.   
 
\subsection{The fixed coupling case}   
  
Let us consider standard leading order evolution of the gluon distribution $xg(x,Q^2)$ 
  
\begin{equation}  
{\partial xg(x,Q^2)\over \partial \ln(Q^2/\Lambda^2)}={\alpha_s\over 2\pi}\int_x^1 {dz\over z}  
P_{gg}(z) \; xg(x/z,Q^2) \; , 
\label{apgf}  
\end{equation}   
where, as usual, the $P_{gg}$ is the gluon-gluon splitting function. 
For simplicity we have neglected possible contribution of the quark distributions.   
In the moment space this equation has the following form 
\begin{equation}  
{\partial g_{\omega}(Q^2)\over \partial \ln(Q^2/\Lambda^2)}={\alpha_s\over 2\pi}  
\gamma_{gg}(\omega)g_{\omega}(Q^2) \; , 
\label{apgfm}  
\end{equation}  
  
where we have defined the Mellin transform to  be 
  
\begin{equation}  
 g_{\omega}(Q^2)=\int_0^1dx\, x^{\omega}  g(x,Q^2) \; ,  
\label{momg}  
\end{equation}  
  
and the gluon anomalous dimension is defined as 
\begin{equation}  
\gamma_{gg}(\omega) \; = \; \int_0^1dz \, z^{\omega} P_{gg}(z) \; . 
\label{momp}  
\end{equation}  
  
The solution of equation (\ref{apgfm}) is straightforward and given by 
  
\begin{equation}  
 g_{\omega}(Q^2)=g_0(\omega)  
\left({Q^2\over Q_0^2}\right)^{{\alpha_s\over 2\pi}\gamma_{gg}(\omega)} \; . 
\label{momgs}  
\end{equation}  
We will now seek equation for the moment function $g_0(\omega)$ using the   
following initial condition 
\begin{equation}  
{\alpha_s\over 2 \pi}xg(x,Q^2=Q_s^2(x))={\alpha_s\over 2 \pi}r^0x^{-\lambda} \; , 
\label{gscal0}  
\end{equation}  
where $Q_s^2(x)$ is given by equation (\ref{qs}). 
The parameter $r^0$ specifies the normalisation of gluon distribution 
along the critical line.    
This boundary condition follows from the geometric scaling condition of the dipole   
proton cross section $\sigma_{dp}(r=1/Q,x)$ which is proportional to   
$\alpha_s xg(x,Q^2) /Q^2$.

In order to find solution for $g_0(\omega)$  we use the inverse Mellin transform   
\begin{equation}  
xg(x,Q^2)={1\over 2 \pi i} \int d\omega\, x^{-\omega} g_{\omega}(Q^2) \; , 
\label{melinv}  
\end{equation}  
where the integration contour should be located to the right of the singularities of   
$g_{\omega}(Q^2)$ in the $\omega$ plane.   
Inserting in equation (\ref{melinv}) the DGLAP solution (\ref{momgs}) for $g_{\omega}(Q^2)$   
we get  
\begin{equation}  
xg(x,Q^2)={1\over 2 \pi i} \int d\omega \,  x^{-\omega} g_0(\omega)  
\left({Q^2\over Q_0^2}\right)^{{\alpha_s\over 2\pi}\gamma_{gg}(\omega)} \; . 
\label{melinv1}  
\end{equation}  
We now set $Q^2=Q_s^2(x)$ with   
the saturation scale $Q_s^2(x)$ defined by equation (\ref{qs}), and require  
the geometric scaling initial condition along the critical line $Q^2=Q_s^2(x)$   
(see eq. (\ref{gscal0})).      
From equations (\ref{qs}),(\ref{gscal0}) and (\ref{melinv1}) we get 
\begin{equation}  
{1\over 2 \pi i} \int d\omega g_0(\omega) x^{-\omega-\lambda{\alpha_s\over 2\pi}\gamma_{gg}(\omega)} =  
r^0x^{-\lambda}  \; . 
\label{gseq1}  
\end{equation}  
This equation can be regarded as the equation for the function $g_0(\omega)$, i.e.   
for the moment of the gluon distribution at the ($x$ independent) scale $Q_0^2$.    
In order to solve this equation   we take the moment   
of both sides of equation (\ref{gseq1}), i.e. we integrate both sides of this equation over   
$dx$ for $0<x<1$   
with the weight $x^{\omega_1-1}$ and get   
\begin{equation}  
{1\over 2 \pi i} \int d\omega {g_0(\omega)\over [\omega_1-\omega-\lambda{\alpha_s\over 2\pi}\gamma_{gg}(\omega)]}    
=  
{r^0\over \omega_1-\lambda} \; . 
\label{project}  
\end{equation}  
We now change the integration variables
\begin{equation}   
z=\omega+\lambda{\alpha_s\over 2\pi}\gamma_{gg}(\omega)  \; , 
\label{chv}  
\end{equation}  
which after inversion specifies the function $\omega(z)$.   
Equation (\ref{project}) in the new variable $z$ takes then the following form 
\begin{equation}  
{1\over 2 \pi i} \int dz {d\omega(z)\over dz} {g_0(\omega(z))\over (\omega_1-z)}    
={r^0\over \omega_1-\lambda} \; . 
\label{project0}  
\end{equation}  
We can easily perform the contour integration  in Eq.(\ref{project0}) and get    
\begin{equation}  
{d\omega(z)\over dz} \bigg|_{z=\omega_1} g_0(\omega(z=\omega_1))    
={r^0\over \omega_1-\lambda} \; . 
\label{project1}  
\end{equation}  
We still need to solve this equation for $g_0(\omega)$ and in order to do this we write  
\begin{equation}  
\omega_1=\omega+\lambda{\alpha_s\over 2\pi}\gamma_{gg}(\omega)   \; , 
\label{invb}  
\end{equation}  
and finally from Eq.(\ref{project1}) we obtain 
\begin{equation}  
g_0(\omega)=[1+\lambda{\alpha_s\over 2\pi}  
{d\gamma_{gg}(\omega)\over d\omega}]{r^0\over[\omega+\lambda{\alpha_s\over 2\pi}  
\gamma_{gg}(\omega)   -\lambda]} \; , 
\label{sol}  
\end{equation}  
 which defines the solution for $g_0(\omega)$. \\  
  
In what follows it is convenient to use directly redefined function $\tilde g_0(z)$   
\begin{equation}  
\tilde g_0(z) \equiv {d\omega(z)\over dz} g_0(\omega(z)) \; , 
\label{gtilde0}  
\end{equation}  
where from equation (\ref{project1}) we  see that   
\begin{equation}  
\tilde g_0(z)= {r^0\over z-\lambda}  \; . 
\label{gtildpole}  
\end{equation}   
The solution of the DGLAP equation with the initial condition specified by equation  
(\ref{gscal0}) then reads 
\begin{equation}  
xg(x,Q^2)={1\over 2 \pi i} \int dz x^{-z}\tilde  g_0(z)  
\left({Q^2\over Q_s^2(x)}\right)^{{\alpha_s\over 2\pi}\gamma_{gg}(\omega(z))}  
\label{solap}  
\end{equation}  
where the integration contour is located to the right of the singularities of   
$\tilde g_0(z)$ and of $\omega(z)$.  If the leading singularity is a pole   
of $\tilde g_0(z)$ at $z=\lambda$ then the leading contribution to $xg(x,Q^2)$   
at small $x$   
is given by 
\begin{equation}   
xg(x,Q^2) \simeq r^0 x^{-\lambda}  
\left({Q^2\over Q_s^2(x)}\right)^{{\alpha_s\over 2\pi}\gamma_{gg}(\omega_0)} \; , 
\label{solap1}  
\end{equation}  
where  
\begin{equation} 
\omega_0=\omega(\lambda) \; . 
\label{omega0} 
\end{equation}  
It should be noted that $\omega_0$ defines position of the pole of $g_0(\omega)$.   
In general we have $\omega_0 \le \lambda$. 
From  equation (\ref{solap1}) we get the following leading small $x$  behaviour for  the gluon density 
${\alpha_s\over 2\pi} xg(x,Q^2) / Q^2$ 
\begin{equation}   
{\alpha_s\over 2\pi}{xg(x,Q^2)\over Q^2} \simeq   
{r^0\over Q_0^2} {\alpha_s\over 2\pi}  
 \left({Q^2\over Q_s^2(x)}\right)^{{\alpha_s\over 2\pi}\gamma_{gg}(\omega_0)-1}\; ,  
\label{gscalap}  
\end{equation}  
which respects the geometric scaling i.e. is a function of only one combined variable $Q^2/Q_s^2(x)$.  Violation of this scaling by the contribution  
of the (branch point) singularity of $\omega(z)$ is a non-leading effect at low  
$x$. 
  
The requirement that the pole of $\tilde g^0(z)$ at $z=\lambda$ is the leading   
singularity imposes certain constraints upon $\lambda$.  In general they are difficult   
to be found exactly since the inversion of equation (\ref{chv}) cannot be performed   
analytically when using complete form of $\gamma_{gg}(\omega)$.  Analytic solution   
of equation (\ref{chv}) is however possible in the double logarithmic approximation   
in which $\gamma_{gg}(\omega)=\gamma_{gg}^{DL}(\omega)$, where  
\begin{equation}  
\gamma_{gg}^{DL}(\omega)={2N_c\over \omega} \; , 
\label{gamdl}  
\end{equation}  
is the most singular in $\omega \rightarrow 0$ part of the gluon anomalous dimension $\gamma_{gg}(\omega)$. 
In this approximation we get 
\begin{equation}  
\omega(z)={z+\sqrt{z^2-4\bar \alpha_s\lambda}\over 2} \; , 
\label{omzdl}  
\end{equation}  
where  
\begin{equation}  
\bar \alpha_s={N_c \alpha_s\over \pi} \; . 
\label{balpha}  
\end{equation}  
We also have 
\begin{equation} 
\omega_0={\lambda + \sqrt{\lambda^2-4\bar \alpha_s\lambda}\over 2} \; . 
\label{omdl0} 
\end{equation} 
The condition that the pole of $\tilde g^0(z)$ at $z=\lambda$ is the leading   
singularity, i.e. that it is located to the right of the branch-point singularity   
of $\omega(z)$ at $z=2\sqrt{\bar \alpha_s \lambda}$ gives the following constraint   
upon the parameter $\lambda$ 
\begin{equation}  
\lambda\ge 4\bar \alpha_s  \; . 
\label{stab}  
\end{equation}  
For $\lambda < 4\bar \alpha_s$ the leading singularity is the branch point of   
$\omega(z)$ at $z=2\sqrt{\bar \alpha_s \lambda}$ and the geometric scaling becomes   
violated.  \\ 
 
It may be interesting to confront our results for the fixed coupling with the properties of  
the exact solution of the non-linear Balitsky - Kovchegov equation \cite{GBMS}.   
In this case geometric-scaling holds for $Q^2 \le Q_s^2(x)$ and the non-linear effects  
can be neglected for $Q^2>Q_s^2(x)$.      
The parameter $\lambda$ specifying the critical line is however not an independent quantity   
and depends upon  the (fixed) coupling $\alpha_s$.   In the double logarithmic approximation it is given by  
$\lambda=4\bar \alpha_s$.  It follows from equation (\ref{stab}) that  this is a  
limiting value of the parameter $\lambda$ for the geometric scaling  
to hold asymptotically  in the small $x$ limit and so for $\lambda=4\bar \alpha_s$  
we expect violation of this scaling for $Q^2>Q_s^2(x)$ down to the very small values of $x$ \cite{GBMS}.\\

\subsection{ Running coupling case}   
 We now pass to the more realistic case with the running coupling. 
In this case the evolution equation for the moment function takes the form:   
  
\begin{equation}  
{\partial g_{\omega}(Q^2)\over \partial \ln(Q^2/\Lambda^2)}={\alpha_s(Q^2)\over 2\pi}  
\gamma_{gg}(\omega)g_{\omega}(Q^2) \; ,  
\label{apgrm}  
\end{equation}  
  
where the running coupling in the leading order is given by  
  
\begin{equation}  
{\alpha_s(Q^2)\over 2\pi}={b\over \ln(Q^2/\Lambda^2)}  \; , 
\label{ralpha}  
\end{equation}  
  
with  
  
\begin{equation}  
b={2\over 11-2/3N_f} \; , 
\label{b}  
\end{equation}  
 with $N_f$ being number of flavours. In this section we consider only gluonic channel therefore we set $N_f=0$. 
The solution of equation (\ref{apgrm}) reads 
  
\begin{equation}  
g_{\omega}(Q^2)= g_0(\omega)\left({\ln(Q^2/\Lambda^2)\over \ln(Q_0^2/\Lambda^2)}\right)^{  
b \gamma_{gg}(\omega)} \; . 
\label{smomr}  
\end{equation}  
From the above solution we obtain 
\begin{equation}  
{\alpha_s(Q^2)\over 2\pi}g_{\omega}(Q^2)= {\alpha_s(Q_0^2)\over 2\pi}g_0(\omega)\left({\ln(Q^2/\Lambda^2)\over \ln(Q_0^2/\Lambda^2)}\right)^{  
b \gamma_{gg}(\omega)-1} \; , 
\label{alg}  
\end{equation}  
  
and so  the result for the gluon distribution $xg(x,Q^2)$ in $x$ space reads in this case 
  
\begin{equation}  
{\alpha_s(Q^2)\over 2\pi}xg(x,Q^2)={1\over 2 \pi i} \int d\omega x^{-\omega} f_0(\omega)\left({\ln(Q^2/\Lambda^2)\over   
\ln(Q_0^2/\Lambda^2)}\right)^{  
b \gamma_{gg}(\omega)-1} \; , 
\label{invm}  
\end{equation}  
  
where   
  
\begin{equation}  
f_0(\omega)={\alpha_s(Q_0^2)\over 2\pi}g_0(\omega)  \; . 
\label{rgom}  
\end{equation}  
  
We now impose the geometric scaling condition (\ref{gscal0}) onto this solution to get 
\begin{equation}  
{1\over 2 \pi i} \int d\omega x^{-\omega} f_0(\omega)\left(1+{\lambda \ln(1/x)\over \ln(Q_0^2/\Lambda^2)}\right)^{  
b \gamma_{gg}(\omega)-1}=r^0x^{-\lambda} \; . 
\label{gscalr}  
\end{equation}  
which is an equation for $f_0(\omega)$.  
 Solution of this equation is complicated, i.e. exact solution generates complicated (branch point)   
singularity of $f_0(\omega)$ at $\omega=\lambda$.  The only observation which we can make is   
that it should generate $x^{-\lambda}$ behaviour softened by inverse powers of the   
$\ln(1/x)$. In order to make some insight into what is going on we have to make some   
approximations.  To be precise let us make the approximation by 
setting $\omega=\lambda$ in the argument of   
$\gamma_{gg}(\omega)$ that gives 
\begin{equation}  
{1\over 2 \pi i} \int d\omega x^{-\omega} f_0(\omega)  
\simeq  
\left(1+{\lambda \ln(1/x)\over \ln(Q_0^2/\Lambda^2)}\right)^{-(  
b \gamma_{gg}(\lambda)-1)} \! \! r^0x^{-\lambda} \; . 
\label{gscalra}  
\end{equation}  
  
Making the same  approximation in the inverse Mellin transform (\ref{invm})   
we get  the solution  
\begin{equation}  
{\alpha_s(Q^2)\over 2\pi}xg(x,Q^2) \simeq x^{-\lambda} r^0\left({\ln(Q^2/\Lambda^2)\over   
\ln(Q_0^2/\Lambda^2)}\right)^{  
b \gamma_{gg}(\lambda)-1}\left(1+{\lambda \ln(1/x)\over \ln(Q_0^2/\Lambda^2)}\right)^{-(  
b \gamma_{gg}(\lambda)-1)}  \, . 
\label{invma}  
\end{equation}  
  
Multiplying and dividing $Q^2$ by $Q_0^2 x^{-\lambda}$   
we finally obtain  
\begin{equation}  
{\alpha_s(Q^2)\over 2\pi}xg(x,Q^2)= x^{-\lambda} r^0  
\left({\ln(Q^2x^{\lambda}/Q_0^2)\over   
 \ln(Q_0^2/\Lambda^2) + \lambda \ln(1/x) }+1\right)^{  
b \gamma_{gg}(\lambda)-1} \; . 
\label{gsviol}  
\end{equation}  
The factor proportional to $\ln(1/x)$ in the denominator of the expression on the r.h.s. of (\ref{gsviol}) generates violation of the geometric scaling. 
Thus in the case of  running of the coupling $\alpha_s(Q^2)$ the scaling behaviour gets violated, it is possible however to factor out the effect of this violation. 
We can also rewrite Eq.(\ref{gsviol}) by using the definition of the saturation scale and the running coupling to get 
\be 
{\alpha_s(Q^2)\over 2\pi} { xg(x,Q^2) \over Q^2} \; = \;   {r^0 \over Q_0^2} {Q_s^2(x) \over Q^2} \left[ 1 + \frac{\alpha_s(Q_s^2(x))}{2 \pi b} \ln(Q^2/Q_s^2(x)) \right]^{b \gamma_{gg}(\lambda)-1} \; , 
\label{gsviol1} 
\ee 
where we see that the violation is proportional to the value of the running coupling evaluated at the saturation scale. 
Consequently when $x \ll 1$ that is when $Q_s(x) \gg 1$ the geometric scaling is restored, 
 provided of course that $ \alpha_s(Q_s^2(x)) \ln(Q^2/Q_s^2(x)) \ll 1$ as well. This condition is equivalent to $\ln(Q^2/Q_s^2(x)) \ll \ln(Q_s^2(x)/\Lambda^2)$.   The same condition  
defining the region in which the geometric scaling holds above the saturation 
scale has recently been found in ref.  \cite{IILMCL}.   
 
\section{Numerical results}  
  
In this section we present numerical results for the evolution of ordinary DGLAP equations  
for the integrated gluon distribution function with special boundary conditions set on the critical line $Q_s^2(x)$ as described in Sec.1.   
\subsection{Fixed coupling case}  
\begin{figure}[t]    
  \vspace*{0.0cm}    
     \centerline{    
         \epsfig{figure=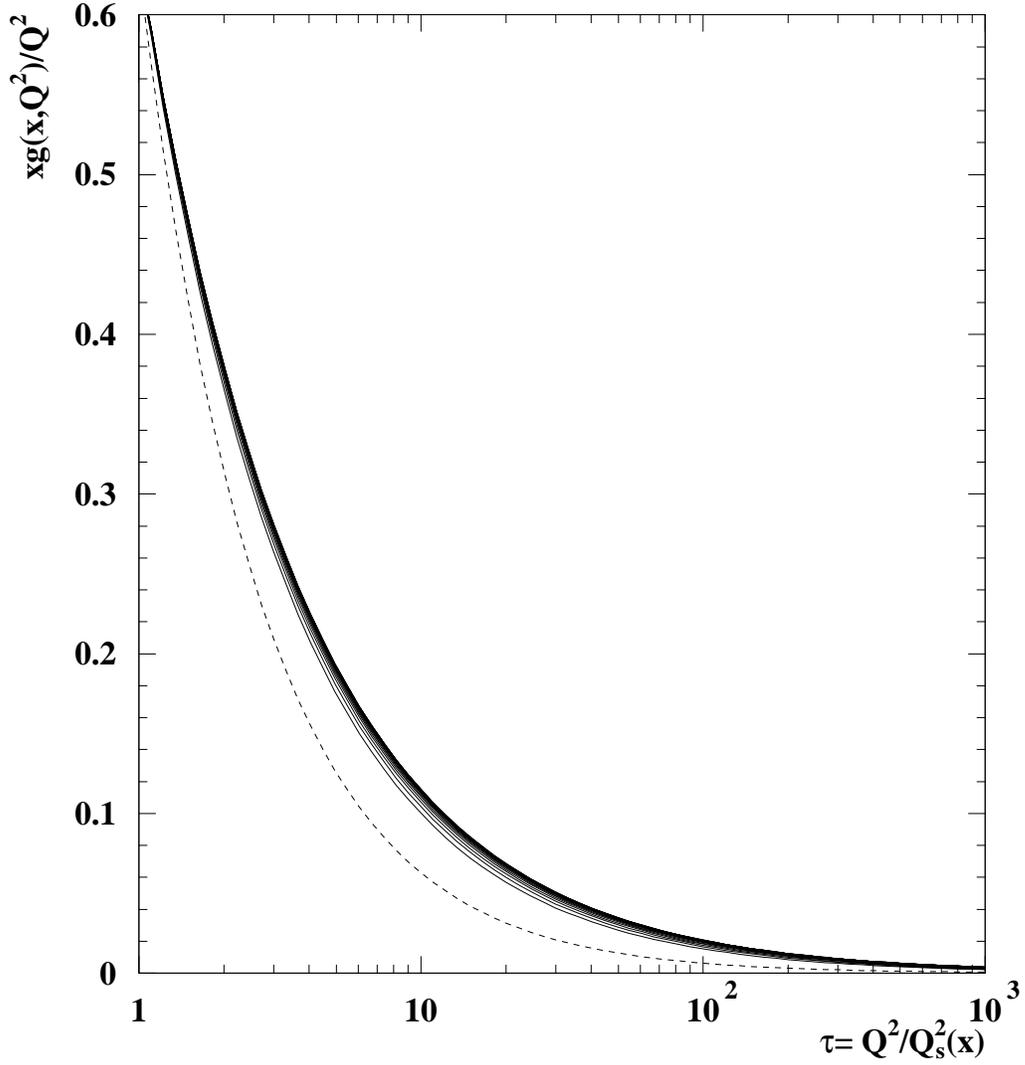,width=15cm}    
           }    
\vspace*{0.5cm}    
\caption{\it Function $xg(x,Q^2)/Q^2$ in DLLA fixed coupling case plotted versus scaling variable  
$\tau = Q^2/Q_s^2(x)$ for different values of rapidities $Y=\ln 1/x$,  
from $Y_{min}=6.0$ to $Y_{max}=46.0$ (solid curves from lower to upper) in steps $\Delta Y = 2$ . 
Dashed curve is the input distribution $\sim 1/\tau$. 
\label{fig:2}}    
\end{figure}

We start with the simplest case which is the fixed strong coupling.  
We assume also in the first approximation the DLLA limit that is we only  
keep the most singular part of the $P_{gg}$ splitting function in our simulation  
i.e.  
\be  
P_{gg}(z) = \frac{2 N_c}{z}, \; \; N_c = 3 \; , 
\label{dlla}  
\ee  
which results in the following form for the anomalous dimension of Eq.(\ref{momp}) 
\be  
\gamma_{gg}(\omega) = \frac{2 N_c}{\omega} \; . 
\label{anomdlla}  
\ee  
The initial condition for the evolution of the gluon density is assumed to be of the form  
(\ref{gscal0}). We take $\lambda = 0.5$ and $\alpha_s = 0.1$.  
In Fig.\ref{fig:2}  
 we show the results of the calculation in this case. We illustrate the scaling behaviour of the gluon density by plotting $xg(x,Q^2)/Q^2$ versus scaling variable $\tau = Q^2 /Q_s^2(x)$ for different values of rapidity $Y=\ln 1/x$.  
 From Eq.(\ref{gscalap}) we see that this function  
should scale with $\tau=Q^2 /Q_s^2(x) $.  
The geometric scaling would correspond in this plot (Fig.~\ref{fig:2}) to the perfect overlap of all curves for different values of $Y$, so that they would form one single line. 
We see that up to a good accuracy  
this function does not depend dramatically on $Y$ and thus on $x$. We do however  
observe that there is some violation of the scaling at large $x$.  This is due to the fact  
that the geometric scaling expression defined by equation (\ref{gscalap}) is only  
expected to hold  
asymptotically in the small $x$ limit.  At finite $x$ this leading behaviour is perturbed by the  
non-leading contribution given by the branch-point singularity of $\omega(z)$ at  
$z=2\sqrt{\bar \alpha_s}$ , (see eq. (\ref{omzdl}) ).\\

\begin{figure}[t]    
  \vspace*{0.0cm}    
     \centerline{    
         \epsfig{figure=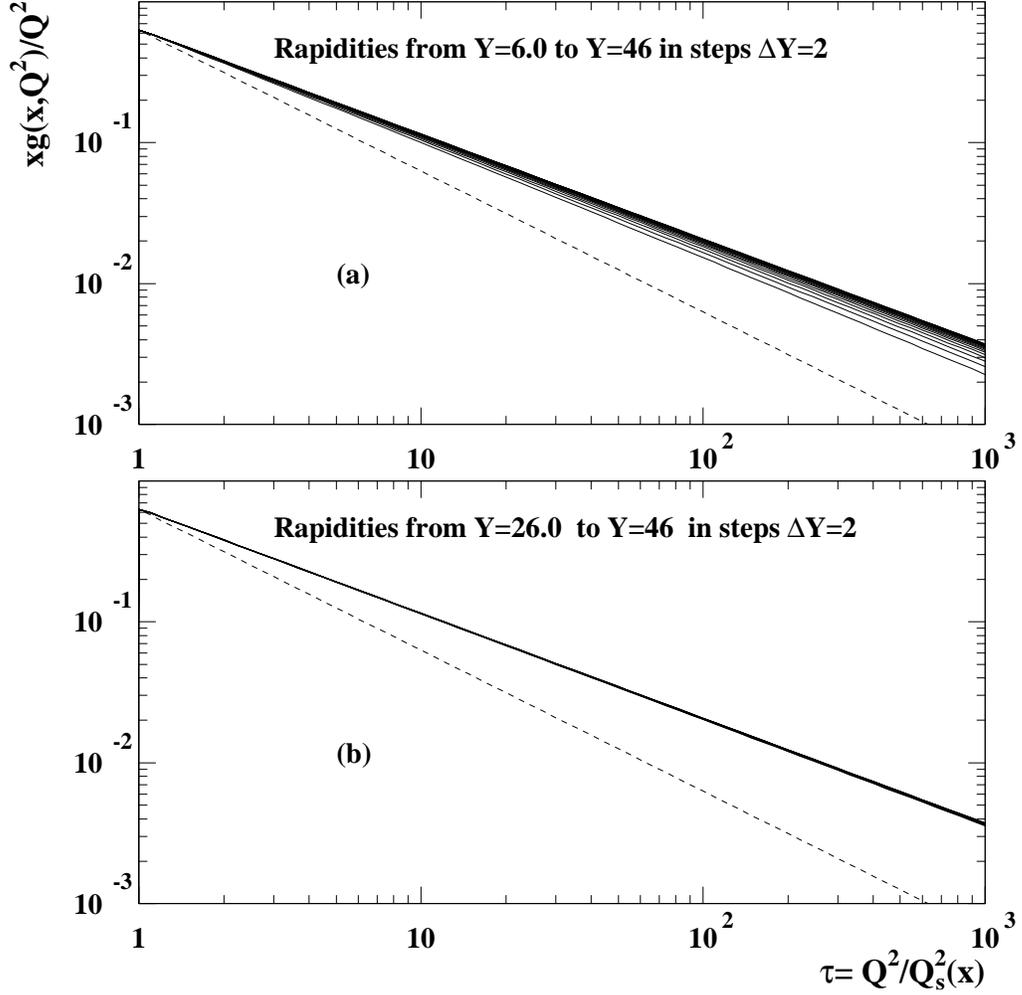,width=15cm}    
           }    
\vspace*{0.5cm}    
\caption{\it Function $xg(x,Q^2)/Q^2$ in DLLA fixed coupling case plotted versus scaling variable  
$\tau = Q^2/Q_s^2(x)$ for different values of rapidities $Y=\ln 1/x$. Solid curves - solutions, dashed curve - input distribution $\sim 1/\tau$.
On upper plot (a): solid curves from lower to upper are for $Y$ rapidities ranging 
from $Y_{min}=6.0$ to $Y_{max}=46.0$ in steps $\Delta Y = 2$.   
Lower  plot (b):  rapidities ranging 
from $Y_{min}=26.0$ to $Y_{max}=46.0$ in steps $\Delta Y = 2$.  
\label{fig:3}}    
\end{figure}

To illustrate better the scaling and its violation we have plotted $xg(x,Q^2)/Q^2$ versus scaling variable $\tau = Q^2 /Q_s^2(x)$ using double-logarithmic scale, see Fig.~\ref{fig:3}a.  
One clearly sees that with increasing rapidity $Y$ the curves do not change  
and reach asymptotic straight line.  
 We have also selected the very low $x$ range   
of Fig.~\ref{fig:3}a, which is Fig.~\ref{fig:3}b. One can see that in this case  
the geometric scaling is nearly preserved (we see nearly single line for different rapidities).   
  
The behaviour of $xg(x,Q^2)/Q^2$ versus $\tau=Q^2/Q_s^2(x)$ is clearly  
governed by a power law, with a power which we estimated to  be approximately  
$-0.77$. From equation (\ref{gscalap}) and (\ref{omdl0}), and using the values 
 of $\lambda$  
and  $\alpha_s$ quoted above we get that the power should be $\frac{\alpha_s}{2\pi}\gamma_{gg}(\omega_0) - 1  = -0.74$  
which is in a very good  agreement with numerical result.  
  
Let us note that in the case of DLLA (\ref{anomdlla}), $\omega_0$ is a solution  of the quadratic equation and is given by (\ref{omdl0}). 
As previously noticed the real solution exists 
 only for $\lambda \ge \lambda_{min} = 4 \bar{\alpha_s}$ with $\bar{\alpha}_s = \alpha_s N_c / \pi$.  
We have numerically checked that for $\lambda \le \lambda_{min}$ our solution no longer exhibits geometric scaling.  
 It is interesting to note, as we have already observed at the end of Sec. 2.1,   that  exactly the same value of $\lambda=\lambda_{min}$ for a power of saturation scale was obtained   
from the studies of the nonlinear Balitsky-Kovchegov equation  
\cite{BAL,KOV1} 
 performed in   
\cite{LEVLUB}  - \cite{GBMS}.    \\

We next abandon the DLLA approximation and consider more general case with  
the full gluon-gluon splitting function $P_{gg}$ which gives the following  
anomalous dimension  
\be  
\gamma_{gg}(\omega) \; = \; 2 N_c \left[ \frac{1}{\omega} - \frac{1}{\omega+1} + \frac{1}{\omega+2} - \frac{1}{\omega+3} - \gamma_E + \frac{11}{12} - \psi(\omega+2) \right] \; , 
\label{anomfull}  
\ee  
where $\psi$ is Polygamma function. In this case  equation (\ref{chv})  
with $z=\lambda$  
can no longer be solved analytically and has to be analysed numerically.  
However, one can get insight into the allowed values of $\lambda$ by making  
the expansion of the anomalous dimension around $\omega=0$.  
In this case $\gamma_{gg}(\omega)/(2 N_c) \simeq \frac{1}{\omega} + A_1(0) + {\cal O}(\omega)$ where $A_1(0) =  - \frac{11}{12}$. Using this approximation in (\ref{chv}) one finds that now geometric scaling will hold if the following condition is satisfied  
\be  
\lambda \ge \lambda_{min} = \frac{4 \bar{\alpha_s}}{[ 1-\bar{\alpha_s}A_1(0) ]^2}  \; . 
\label{lamcritfull} 
\ee  
We have checked numerically that above approximation works very well and gives  
very close results to the solution of (\ref{chv}) with full $\omega$ dependence of anomalous dimension $\gamma_{gg}(\omega)$.  
  
\begin{figure}[t]    
  \vspace*{0.0cm}    
     \centerline{    
         \epsfig{figure=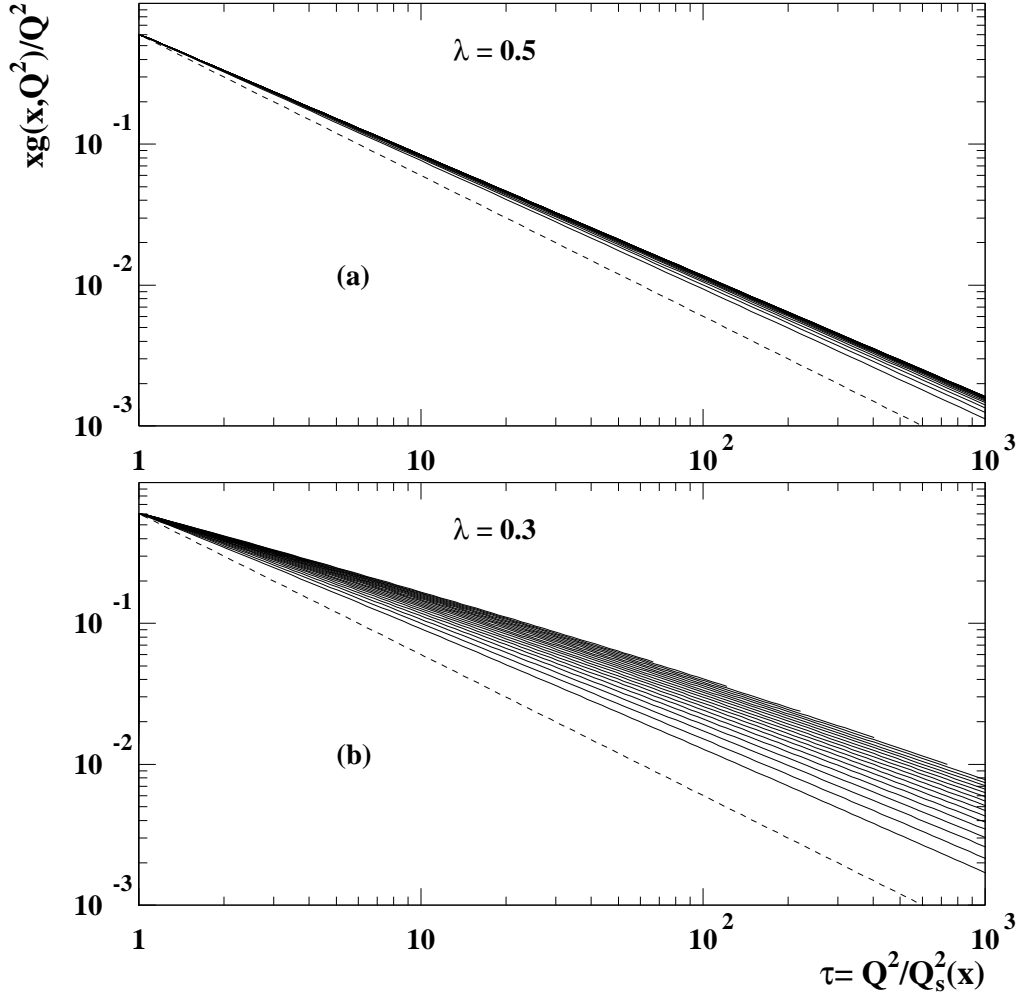,width=15cm}    
           }    
\vspace*{0.5cm}    
\caption{\it Fixed coupling case with complete gluon anomalous dimension $\gamma_{gg}(\omega)$. 
 Function $xg(x,Q^2)/Q^2$ plotted versus scaling variable  
$\tau = Q^2/Q_s^2(x)$ for different values of rapidities $Y=\ln 1/x$  
from $Y_{min}=6.0$ to $Y_{max}=46.0$ in steps $\Delta Y = 2$.    
Upper plot (a): scaling exponent $\lambda=0.5$, lower plot (b): scaling exponent $\lambda=0.3$ .  
\label{fig:4}}    
\end{figure}

In Fig.\ref{fig:4}a we plot $xg(x,Q^2)/Q^2$ as a function of scaling  
variable $Q^2/Q_s^2(x)$ in the case of calculation with the full anomalous dimension (\ref{anomfull}). We have taken $\lambda=0.5$, and $\bar{\alpha}_s=0.1$. We see that the function exhibits geometric scaling (although there is some residual violation at larger values of $x$). The calculated value of exponent from numerical calculation is $-0.85$ which is again in nearly perfect agreement with  the analytical estimate based on the approximation described above which gives $-0.86$.  
We also present in Fig.~\ref{fig:4}b the calculation in the case of $\lambda=0.3$  
which is below the critical value (\ref{lamcritfull}) equal in this case $\lambda_{min}=0.33$ for   
$\bar{\alpha}_s=0.1$. We clearly see that the geometric scaling is never present in that case.   

One can study the scaling and its violation in a more quantitative way by examining the following expression
\be
\label{derivative}
\Delta(Y,\tau) \; = \; \frac{1}{h} \frac{\partial h(Y,\tau)}{\partial Y} \bigg |_{\tau={\rm fixed}} \; ,
\ee
where
\be
\label{defh}
h(Y,\tau) \equiv \frac{\asb}{Q^2} xg(x,Q^2) \; .
\ee
Derivative $\Delta(Y,\tau)$ should vanish in the region where geometric scaling is satisfied. Consequently its deviation from zero will characterise the scaling violation of solution (\ref{defh}).

\begin{figure}[t]    
  \vspace*{0.0cm}    
     \centerline{    
         \epsfig{figure=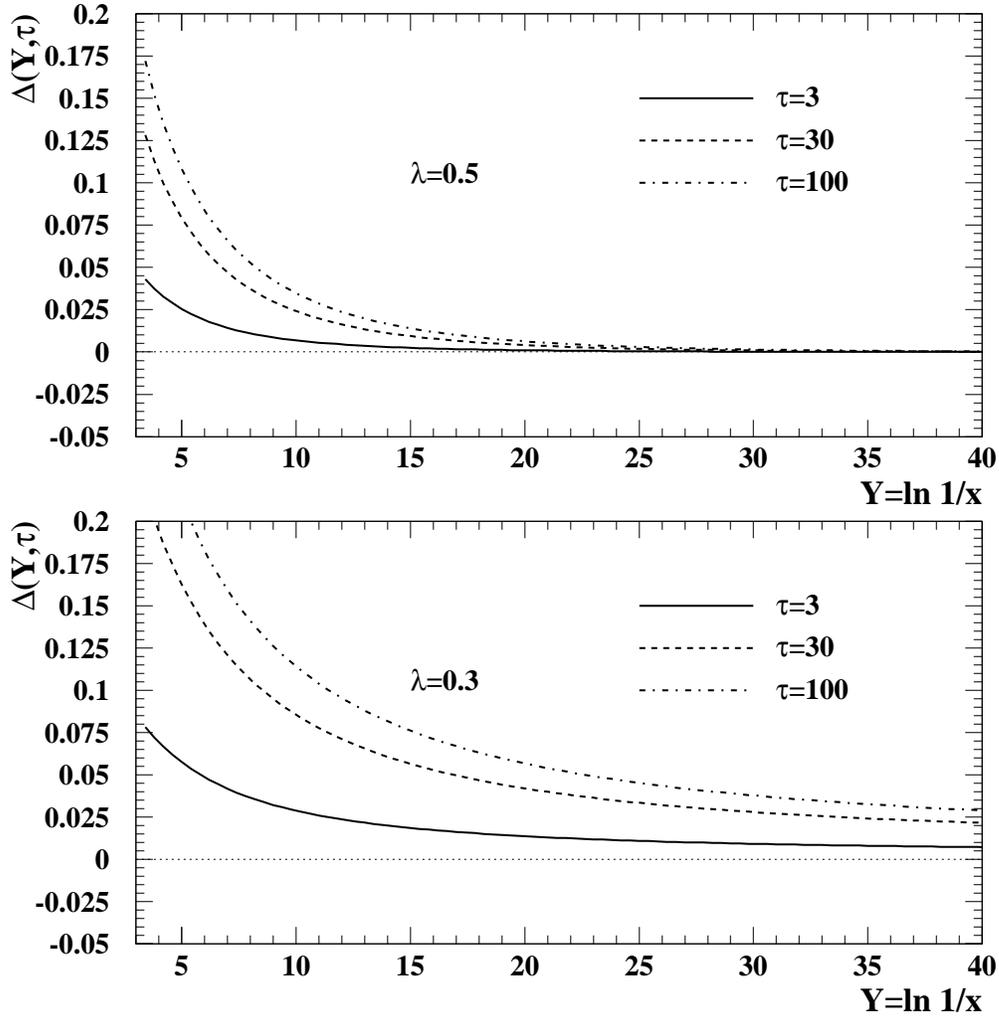,width=15cm}    
           }    
\vspace*{0.5cm}    
\caption{\it  The derivative $\Delta(Y,\tau)$ from 
Eq.(\ref{derivative}) as a function of rapidity $Y$ for various values of the scaling variable $\tau$.
Upper plot - scaling exponent $\lambda=0.5$, lower plot - scaling exponent $\lambda=0.3$.
\label{fig:5}}    
\end{figure}

We present the quantity $\Delta(Y,\tau)$ in Fig.\ref{fig:5} for the case of calculation with complete anomalous dimension and two selected values of $\lambda:$  $0.3$ and $0.5$. Derivative $\Delta(Y,\tau)$ in Fig.\ref{fig:5} therefore illustrates the scaling and its violation for the solution shown in  Fig.\ref{fig:4}. From Fig.\ref{fig:5} it is clear that in $\lambda=0.5$ the scaling is always reached, even for high values of $\tau$ , that is very far right to the critical line.
On the other hand in the case $\lambda=0.3$ the derivative $\Delta(Y,\tau)$  never vanishes meaning that for $\lambda<\lambda_{min}$ the solution of the DGLAP equation with the fixed coupling  case does not exhibit geometric scaling.
 
\subsection{Running coupling case}  
  
\begin{figure}[t]    
  \vspace*{0.0cm}    
     \centerline{    
         \epsfig{figure=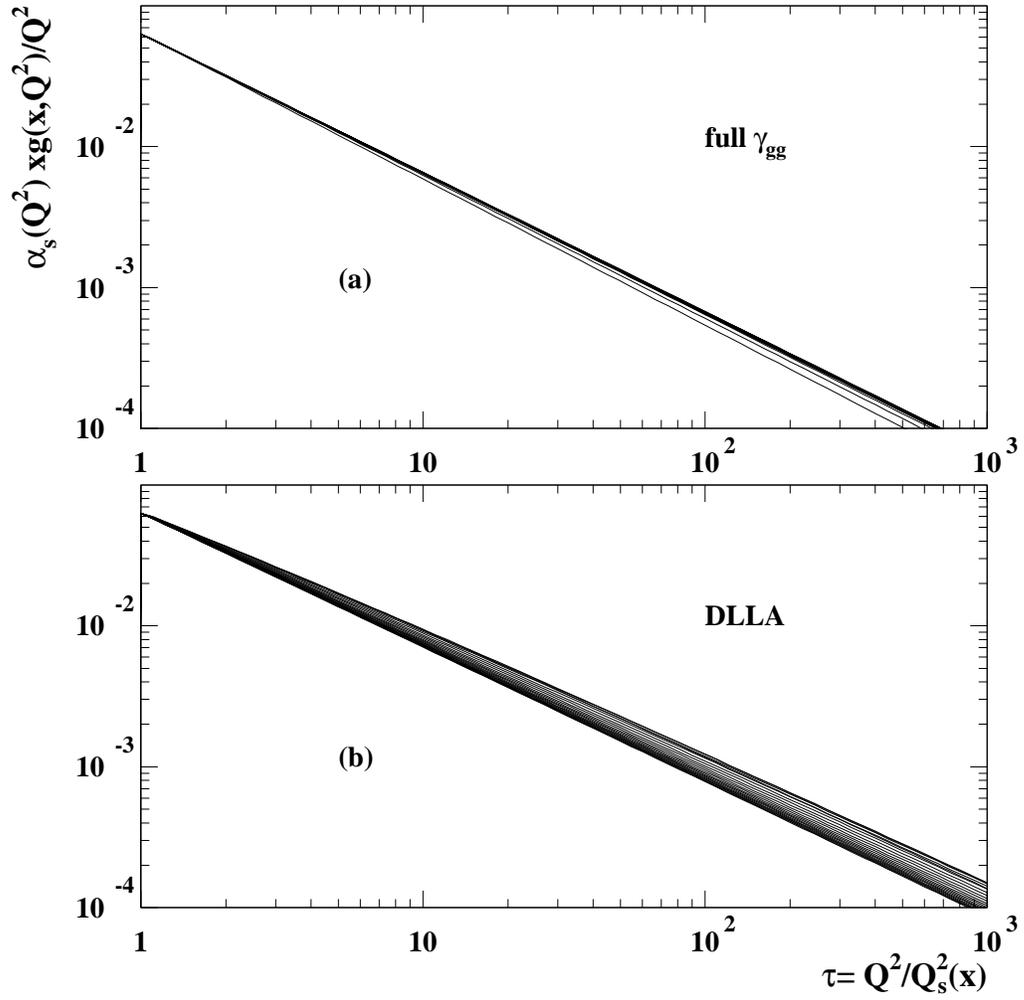,width=15cm}    
           }    
\vspace*{0.5cm}    
\caption{\it   The solution $\alpha_s(Q^2) xg(x,Q^2) / Q_s^2(x) $ in the running coupling case.  
Rapidity range from  $Y_{min}=6.0$ to $Y_{max}=46.0$ in steps $\Delta Y = 2$. 
Upper plot (a): case with full anomalous dimension $\gamma_{gg}(\omega)$, 
 lower plot: (b): case in DLLA approximation. 
\label{fig:6}}    
\end{figure}

We consider now the case in which $\alpha_s$ is running and study the impact of scaling boundary  
condition (\ref{gscal0}) onto the evolution.   
We consider full expression for the anomalous dimension in that case 
$\gamma_{gg}(\omega)$ given by Eq.(\ref{anomfull}). 
The running of the coupling requires that the evolution is taken in the region well above the Landau  
pole. In our case this means that one has to evolve with $Q^2 > Q_s^2(x)$ and we would like to have $Q_s^2(x)$  
big enough for all values of $x$.   
For the purpose of illustration we take that $Q_s^2(x) = Q_0^2 (x/x_0)^{-\lambda}$ where $Q_0^2 = 1.0 \; \rm GeV^2$  
and $x_0 = 1.0$. This means that at $x=1$ the saturation scale is equal $Q_s^2 = 1.0 \; \rm GeV^2$.  
This assumption might seem artificial   
considering the present phenomenology of lepton-nucleon scattering which suggests  
that saturation scale could be of the order of $1 \; \rm GeV^2$ at $x \simeq 10^{-4}$ for the most central  
collisions at HERA collider \cite{KGBMW,MSM}.  
However, we use it here for the purpose of illustration of basic effects  
of the evolution with special scaling boundary conditions. 
We concentrate ourselves here on presenting  general properties of the solution rather than trying to describe the experimental data.  
 We also take $N_f=0$, that is we are considering pure gluonic channel.  
In Fig.\ref{fig:6}a  we present the results of the calculation  by plotting  
$\alpha_s(Q^2) xg(x,Q^2)/Q^2$ versus scaling variable $\tau = Q^2 / Q_s^2(x)$ in the case with full gluon anomalous dimension. 
For comparison we also show the calculation performed  in the DLLA approximation Fig.~\ref{fig:6}b.  
We see that the geometric scaling is mildly violated in the running coupling case, more strongly in the DLLA approximation due 
to the faster evolution.
This fact can be understood on the basis of Eq.(\ref{gsviol1}) where the numerical value of the exponent of expression on the r.h.s. is much bigger in the DLLA case: $b\gamma_{gg}(\lambda=0.5)=0.18$ in the case with full anomalous dimension and  $b\gamma_{gg}(\lambda=0.5)=13/11$ in DLLA case.

\begin{figure}[t]    
  \vspace*{0.0cm}    
     \centerline{    
         \epsfig{figure=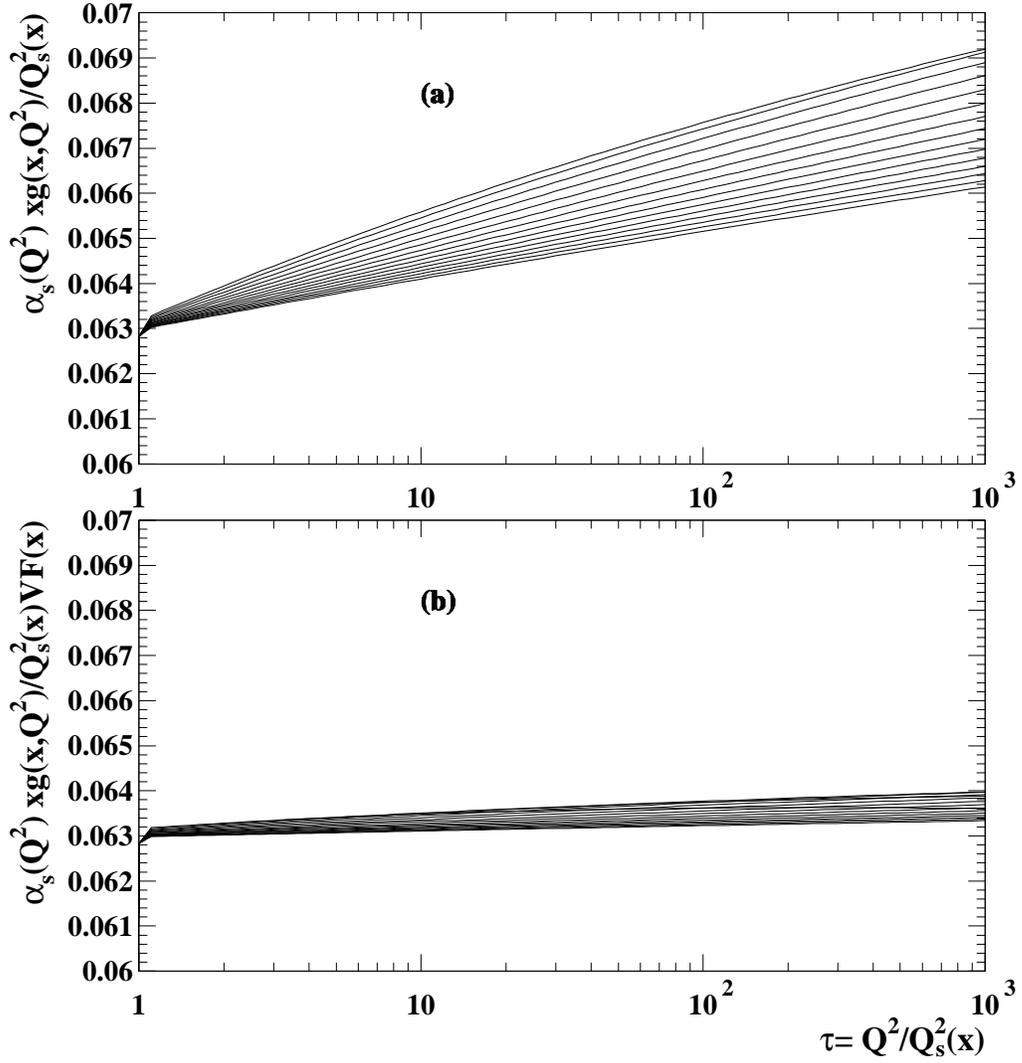,width=15cm}    
           }    
\vspace*{0.5cm}    
\caption{\it   The solution $\alpha_s(Q^2) xg(x,Q^2) / Q_s^2(x)  $ in the running coupling case.  
We have selected high rapidity range from  $Y_{min}=18.0$ to $Y_{max}=46.0$ in steps $\Delta Y = 2$. 
 Upper plot (a): $\alpha_s(Q^2) xg(x,Q^2) / Q_s^2(x)  $, 
lower plot (b): $\alpha_s(Q^2) xg(x,Q^2) / Q_s^2(x) VF(x)  $ where the factor $VF(x)$ is a scaling violation factor  
defined in equation (\ref{gsviolvf}). 
\label{fig:7}}    
\end{figure}

We have tried to estimate whether the violation is consistent with the analytical prediction of formula  
(\ref{gsviol}). In Fig.~\ref{fig:7}a we present the same quantity as in Fig.~\ref{fig:6}a but  
multiplied by the scaling variable $\tau = Q^2 / Q_s^2(x)$. 
The solid black curves in Fig.~\ref{fig:7}a from the upper to lower are for decreasing values of $x$.  
One can see that the solution exhibits some small violation of the geometric scaling and that  
the magnitude of this violation is smaller for smaller values of $x$ ( the curves are becoming  
closer and closer as $x$ decreases ). This is consistent with general behaviour  
predicted by equation  
(\ref{gsviol}) where the scaling violating factor on the r.h.s. tends to unity  
when $\ln(1/x) \gg 1$. We stress that the observed scaling violation is very small in this kinematical regime. For example at very high value of $\tau=10^3$ the violation of the scaling is about $5\%$ 
in a huge rapidity range from $Y=6$ to $Y=46$. \\ 
  
It follows from equations (\ref{gsviol}) and (\ref{gsviol1}) that the violation of the geometric scaling  
can be approximately factored out.   
We    checked this approximate prediction   
by considering the quantity  
\begin{eqnarray} 
& & \alpha_s(Q^2) xg(x,Q^2) / Q_s^2(x) \, VF(x) \nonumber \\ 
& & \nonumber \\ 
{\rm with} & & VF(x) =   \left[\frac{\ln(Q^2/Q_s^2(x) )}{\ln(Q_0^2/\Lambda^2) + \lambda  \ln(1/x) } + 1\right]^{1-b\gamma_{gg}(\lambda)}  
\label{gsviolvf} 
\end{eqnarray} 
which according to equation (\ref{gsviol}) should be constant with respect to $\tau = Q^2/Q_s^2(x)$.  
The results for the above quantity are  shown in Fig.~\ref{fig:7}b (which is Fig.~\ref{fig:7}a multiplied by $VF(x)$ ) where now we see that approximately  
the geometric scaling is nearly restored (curves form a very narrow band) at high values of rapidity .\\  
 
\begin{figure}[t]    
  \vspace*{0.0cm}    
     \centerline{    
         \epsfig{figure=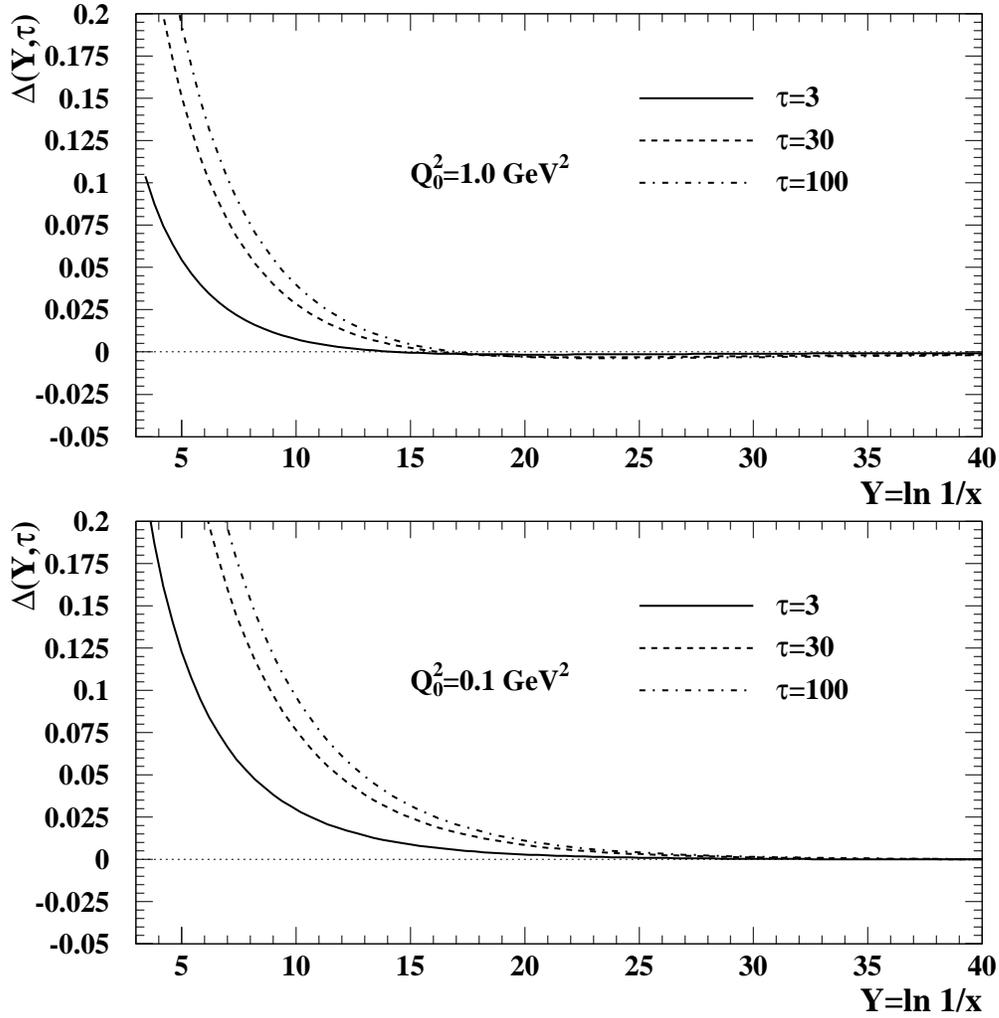,width=15cm}    
           }    
\vspace*{0.5cm}    
\caption{\it The derivative $\Delta(Y,\tau)$ defined in 
Eq.(\ref{derivative}) as a function of rapidity $Y$ for different values of $\tau$  and different choices of normalisation $Q_0^2$. The scaling exponent $\lambda$ was set to be equal $\lambda = 0.5$.  
\label{fig:8}}    
\end{figure}

Also in the case of running coupling we have studied the features of the geometric scaling using the method of the derivative, see Eq. (\ref{derivative}).
The results are shown in Fig.\ref{fig:8} where it is clear that there is always a region where the geometric scaling is (approximately) preserved in the running coupling case, even at very high values of $\tau$. This is consistent with formula (\ref{gsviol1}) provided we have $\asb(Q_s^2) \ln \tau \ll 1, \; x \ll 1$ and also with the conclusions of ref. \cite{IILMCL,AHMDT}. 
We have also illustrated in Fig.\ref{fig:8} the sensitivity of the results on the variation of the normalisation for the saturation scale i.e. $Q_0^2$.
Changing the parameter $Q_0^2$ from $1$ - upper plot in Fig.\ref{fig:8} to $0.1 \; \; {\rm GeV^2} $ - lower plot in Fig.\ref{fig:8}, influences the size of the violation of the scaling. One can see that the geometric scaling is postponed to the higher values of rapidity.
\section{Summary and conclusions} 
 
In this paper we studied effects of the DGLAP evolution upon the geometric scaling.  We solved the DGLAP evolution equation for the gluon distribution with the initial 
 condition respecting the geometric scaling and provided along the critical  
line \mbox{$Q^2=Q_s^2(x)$}.   
In the case of the fixed QCD  coupling we obtained analytic solution of the DGLAP equation  
with those boundary condition, Eq. (\ref{solap}). We also showed that  for sufficiently large values of the  
parameter $\lambda$ defining the critical  line  
  this  solution of the DGLAP equation preserves the  
geometric scaling for the leading term at small $x$, (see Eq. (\ref{solap1})).  In the double logarithmic approximation  
of the DGLAP equation this happens for $\lambda \ge 4 \bar \alpha_s$, where  
$\bar \alpha_s$ is defined by  equation (\ref{balpha}). Geometric scaling is however  
violated by  effects which are  subleading  at small values of $x$.   
We have also obtained approximate solution of the DGLAP equation with the running  
coupling starting again from the boundary conditions respecting geometric scaling  
 along the critical line.     
In the running  coupling case geometric scaling is mildly violated for arbitrary values of the  
parameter $\lambda$ yet this violation can be approximately  factored out.   
The size of this small violation is controlled by the quantity $\asb(Q_s^2) \ln Q^2/Q_s^2$.
Thus in the region where $x \ll 1$ and $\ln Q^2/Q_s^2 \ll \ln Q_s^2/\Lambda^2$ the geometric
scaling in the running coupling case is preserved.
Results of the detailed numerical analysis confirmed all those expectations.\\ 
 We conclude that  the geometric scaling is a very useful regularity following from  
the saturation model.   
We believe that it might be interesting to incorporate  this  'DGLAP improved' geometric scaling  
in the phenomenological  
analysis of the data.              
 
\section*{Acknowledgments} 
This research was partially supported 
by the EU Fourth Framework Programme `Training and Mobility of Researchers', 
Network `Quantum Chromodynamics and the Deep Structure of Elementary 
Particles', contract FMRX--CT98--0194 and  by the Polish  
Committee for Scientific Research (KBN) grants no. 2P03B 05119, 2P03B 12019 and 5P03B 14420.

\newpage  
 
\newpage    
\newpage    

\newpage    
\newpage    
\newpage    
\newpage    

\end{document}